\documentclass[rnote,traditabstract]{aa} 
\usepackage{subfigure}
\usepackage{longtable}
\usepackage{graphicx}
\usepackage{graphics}
\usepackage{keyval}
\usepackage{trig}
\usepackage{psfig,epsf}
\usepackage{txfonts}
\def\beq{\begin{equation}}
\def\eeq{\end{equation}}

\begin{document}

\title{Isolated and non-isolated dwarfs in terms of modified Newtonian
dynamics}

\author{G. Gentile\inst{1} \and G. W. Angus\inst{2} \and
  B. Famaey\inst{3,4} \and S.-H. Oh\inst{5,6} \and W. J. G. de
  Blok\inst{7}}

\institute{
$^{1}$ Sterrenkundig Observatorium, Universiteit Gent, Krijgslaan
  281-S9, B-9000 Gent, Belgium\\
$^{2}$ Astrophysics, Cosmology \& Gravity Centre, University of Cape
Town, Private Bag X3, Rondebosch, 7700, South Africa \\
$^{3}$ Observatoire Astronomique, Universit\'e de Strasbourg, CNRS UMR 7550,
  F-67000 Strasbourg, France \\
$^{4}$ AIfA, Universt\"at Bonn, D-53121 Bonn, Germany \\
$^{5}$ International Centre for Radio Astronomy Research (ICRAR),
Univ. of Western Australia, 35 Stirling Highway, Perth, WA 6009, AU \\
$^{6}$ ARC Centre of Excellence for All-sky Astrophysics (CAASTRO) \\
$^{7}$ ASTRON, the Netherlands Institute for Radio Astronomy, Postbus 2, 7990 AA, Dwingeloo, The Netherlands
}

\abstract{
Within the framework of modified Newtonian dynamics (MOND)
we investigate the kinematics of two
dwarf spiral galaxies 
belonging to very different environments, namely KK~246
in the Local Void and Holmberg II in the M81 group. A mass model of
the rotation curve of KK~246 is presented for the first time, and we
show that its observed kinematics are consistent with MOND. 
 We re-derive the outer rotation curve of Holmberg II, by modelling its HI data cube,
and find that its inclination should be closer to face-on than previously
derived. This implies that Holmberg II has a higher rotation velocity in its outer
parts, which, although not very precisely constrained, is 
consistent with the MOND prediction. 
}


\maketitle

\section{Introduction}

One of the biggest challenges facing the current Lambda cold dark
matter (LambdaCDM) model of
cosmology is the observational appearance of an acceleration constant
$a_0 \sim 10^{-10} {\rm m~} {\rm s}^{-2} \sim c \sqrt{\Lambda}$  in
many apparently unrelated scaling relations between dark matter and
baryons in galaxies (see, e.g., McGaugh 2004, Donato et al. 2009,
Gentile et al. 2009, Famaey \& McGaugh 2012). These scaling relations,
including the baryonic Tully-Fisher relation, the Freeman limit for
stable pure disks, the universality of dark and baryonic surface
densities within the halo core radius, and more generally the mass
discrepancy-acceleration relation, might all require a large 
amount of fine-tuning for collisionless dark matter
models. Surprisingly, all these relations can all be summarized by the empirical
formula of Milgrom~(1983), which the modified Newtonian
dynamics (MOND) paradigm is based on, linking the true gravitational attraction
$g$ to the Newtonian gravitational field $g_N$ (calculated from the
observed distribution of visible matter) by $g=(g_N a_0 )^{1/2}$ in
the limit of $g_N \ll a_0$. The success of this formula would mean
that the observed gravitational field in galaxies mimicks a
universal force law generated by the baryons alone. In the case of inert,
collisionless, and dissipationless dark matter, Milgrom's law
would probably precisely emerge only after an unreasonable amount of
fine-tuning in the expected feedback from the baryons. The
relation between the distribution of baryons and dark matter should indeed
depend on the various different histories of formation, intrinsic
evolution, and interaction with the environment of the various
different galaxies, whereas Milgrom's law provides a successful, unique,
environment-independent, and history-independent relation. This is the
strongest argument to consider MOND as a serious alternative to the
current cosmological model.

\begin{figure*}
\begin{center}
\includegraphics[width=0.9\textwidth]{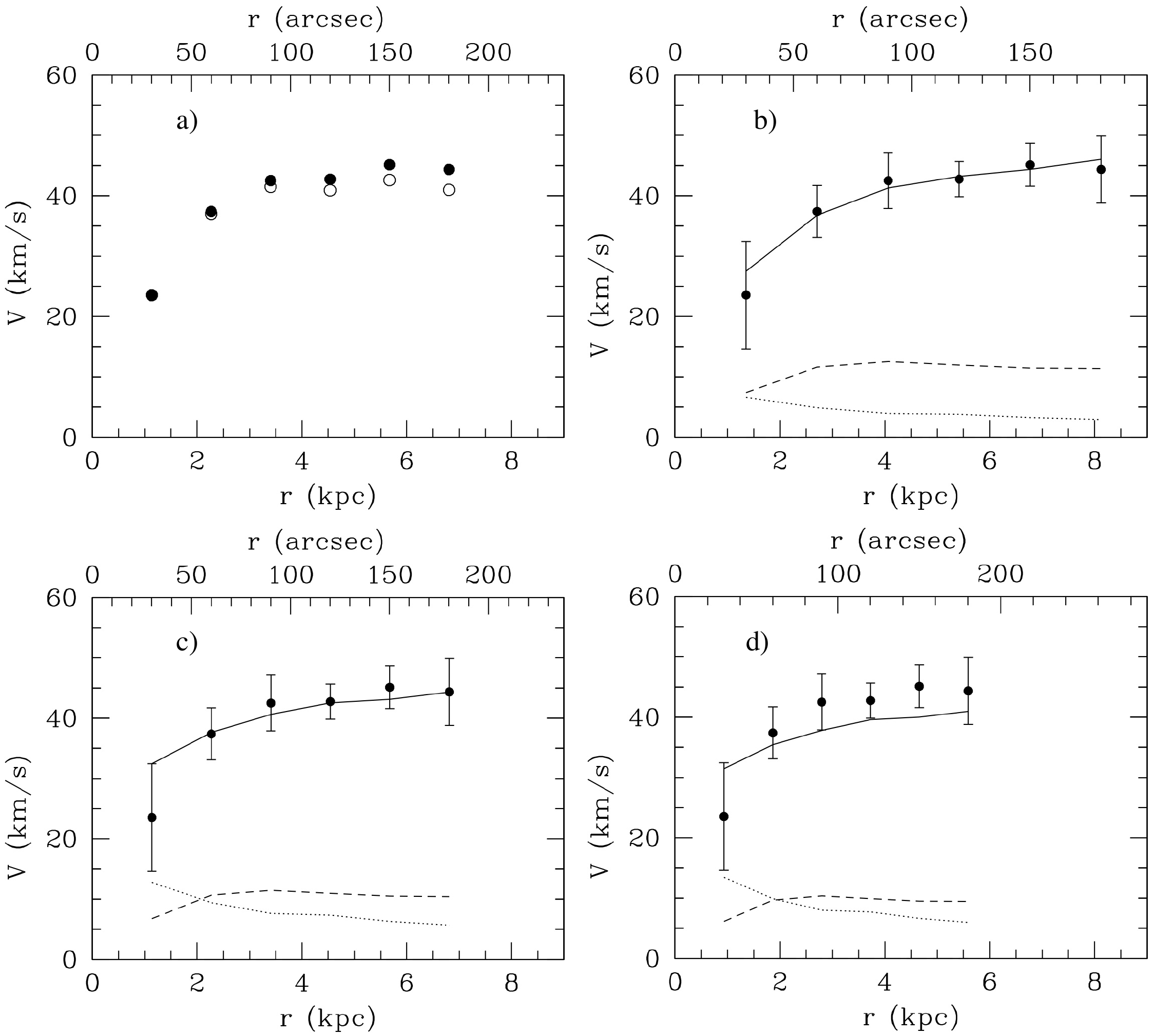}
\end{center}
\caption{
All the panels are based on the rotation curve of KK~246 from Kreckel et al. (2011),
corrected for beam-smearing (see text for details).
Panel a): Asymmetric-drift corrected (full circles) and uncorrected
  (open circles) rotation curves of KK~246, assuming a distance
    of 7.8 Mpc. 
Panel b), c), and d): MOND fits of the rotation curve with distances of 9.3 Mpc
  (the best-fit value, which is only weakly constrained), 7.8 Mpc
  (Karachentsev et al. 2006), and 6.4 Mpc
  (Tully et al. 2008), respectively. The best-fit stellar M/L ratios
  (in the H-band) are 0.22, 0.98, and 1.32, respectively. 
The dotted line represents the Newtonian stellar contribution to the rotation
curve, the dashed line is the Newtonian gas disk contribution, and the solid
line is the total MOND fit.
}
\label{rc_all}
\end{figure*}

It is thus a fundamental prediction of MOND (and, if verified, a
challenge to $\Lambda$CDM) that all galaxies should obey Milgrom's
law\footnote{Modulo a possible effect of a strong gravitational field
  environment, known as the external field effect, and with some
  dependence on the assumed ``interpolating function" between the
  Newtonian regime $a_N$ and MONDian regime  $g=(g_N a_0 )^{1/2}$
  (see, e.g., Famaey \& McGaugh 2012).} independently of their history
and environment, regardless of whether they reside in large voids or dense
groups. For instance, KK~246 and Holmberg~II are two dwarf irregular
galaxies in very different environments: KK~246 is a very isolated
galaxy in the Local Void (Tully et al. 2008), while Holmberg~II
belongs to the M81 group of galaxies
(e.g. Karachentsev 2005, Walter et al. 2007). If MOND is correct, they
should both conform to Milgrom's law. 
However, 
the recently determined rotation curve of Holmberg II (Oh et al. 2011)
 seems a priori inconsistent with MOND, because Milgrom's law severely
 overpredicts the observed rotation curve. 
The isolated galaxy KK~246, on the other hand, has a maximum velocity
consistent with MOND (see also Milgrom 2011), 
despite having a very high mass discrepancy: its
dynamical-to-baryonic mass ratio is $\approx$15. We
note that despite its many successes in terms of global scaling
relations and
dozens of individual galaxy rotation curves, a few galaxies indeed
remain challenging for MOND (for instance NGC 3198, see Gentile,
Famaey \& de Blok 2011), but none as much as the published rotation
curve of Holmberg~II.

Here, we re-analyse in detail the MOND rotation curves of these two
galaxies. The previous analyses of KK~246 are not as detailed as the
one presented here, because the whole set of data necessary to perform
a mass 
decomposition based on the rotation curve was unavailable. On the
other hand, we investigate the inclination angle of Holmberg~II in
detail, building model data cubes and
comparing them to the observed ones, thereby showing that the
previously derived inclination angle was too high, at least in the
outer parts (as could also be
qualitatively inferred from the axis ratio in the total HI map).

\section{KK~246}

KK~246 is a very isolated galaxy in the Local
Void (Tully et al. 2008); its distance was estimated to be 7.8 $\pm$
0.6 Mpc by
Karachentsev et al. (2006) using the tip of the red giant branch (TRGB)
method. However, Tully et
al. (2008) claimed that, because of the reddening estimate they used,
Karachentsev et al.'s might have been too high. This new reddening value 
brings the distance down by almost 20\%, to 6.4 Mpc. 
We assumed a distance of 7.8 Mpc, but 
we also assumed an uncertainty on the distance of 20\%, to include intrinsic
uncertainties in the TRGB method and reddening uncertainties.

Atomic hydrogen HI data of KK~246 were presented by Kreckel et
al. (2011), who analysed
Very Large Array (VLA) and Expanded Very Large Array (EVLA) 
data. Amongst other results, they derived a rotation curve and a neutral
hydrogen surface
density profile. The rotation curve was derived
based on tilted-ring modelling of the intensity-weighted velocity field,
which is probably not the optimal way to derive the rotation curve
because the velocity field is quite poorly resolved. In addition, the
uncertainties in the published rotation curve are likely to be
unrealistically small, as they range between 0.19 km s$^{-1}$ and 0.75 km
s$^{-1}$. Following Sicking (1997), 
to account for the correlation of the data,
the formal uncertainties in the
tilted-ring fit were multiplied by a factor $\sqrt{4 \pi B_{0x}
  B_{0y}/\delta x \delta y} \sim 12$, where $B_{0x}$ and $B_{0y}$ describe the
size of the Gaussian
beam, exp$\left(-\left(\frac{x^2}{2B_{0x}^2}+\frac{y^2}{2B_{0y}^2}\right)\right)$, and
$\delta x$ and $\delta y$ are the pixel size.

Since the
velocity field is poorly resolved and the rotation curve was derived
from the intensity-weighted mean velocity field, the innermost parts
of the rotation curve are likely to be underestimated due to 
beam-smearing, as also qualitatively visible in the channel maps and
position-velocity diagram published by Kreckel et al. (2011). 
We estimated the beam-smearing correction to the Kreckel et 
al. rotation curve by making use of the similarity between
KK~246 and NGC 3741 in terms of the inclination angle and the inner rotation
curve shape and amplitude. For the latter galaxy, Begum et al. (2005) 
derived various rotation curves by making tilted-ring models of the
intensity-weighted mean velocity fields at different angular
resolutions. The velocity difference between the rotation curve at the
highest angular resolution and the rotation curve at the resolution
comparable to the Kreckel et al. (2011) observations gave us an
estimate of the beam-smearing correction, which we found to decrease
from $\sim$7 km s$^{-1}$ for the innermost point to $\sim$3 km
s$^{-1}$ at a radius of 90 arcsec. Keeping in mind that this is only a
rough estimate, we performed the rest of the analysis on the rotation curve
corrected for beam-smearing.

In small dwarf spiral galaxies such as KK~246, the rotation velocity is
only a factor of a few times larger than the gas velocity
dispersion. This means that the observed rotation velocity 
$V_{\rm rot}$ has to be
corrected for pressure support (asymmetric drift). 
Following Bureau
\& Carignan (2002), we derived the corrected rotation velocity $V_{\rm cor}$ using

\begin{equation}
V_{\rm cor}^2 = V_{\rm rot}^2 + \sigma_{\rm D}^2,
\end{equation}

\noindent
where $\sigma_{\rm D}$ is the asymmetric drift correction, derived
from

\begin{equation}
\sigma_{\rm D}^2 = -R \sigma^2 \frac{\partial {\rm ln}(\rho
  \sigma^2)}{\partial R} = -R \sigma^2 \frac{\partial {\rm ln}(\Sigma
  \sigma^2)}{\partial R},
\label{sigmad}
\end{equation}

\noindent
assuming a constant scale-height. Here $\sigma$ is the velocity dispersion, $\Sigma$ is the surface
density (from Kreckel et al. 2011), and $\rho$ is the volume density.

The
derivative in eq. \ref{sigmad} can have large fluctuations, thus
following Oh et al. (2011) we 
fitted the product $\Sigma \sigma^2$ with the function

\begin{equation}
\Sigma \sigma^2 = \frac{I_0 (R_0 + 1)} {R_0 + e^{\alpha R}}
\end{equation}

\noindent
where $I_0$, $R_0$, and $\alpha$ are parameters, and we assumed a
constant $\sigma$ of 8 km s$^{-1}$. We found that the
maximum correction is about 3 km s$^{-1}$, as shown in
Fig. \ref{rc_all}.

To make mass models based on the rotation curve, photometric
data were also needed to model the contribution of the stellar disk. We
used the data of Kirby et al. (2008), who presented deep H-band images
of a sample of nearby galaxies (including KK~246), and the
corresponding surface brightness profiles.

The asymmetric-drift corrected rotation curve $V_{\rm cor}$ was decomposed into its
stellar and gaseous components, $V_{\rm stars}$ and $V_{\rm gas}$
respectively. We adopted the ``simple'' MOND interpolating function
(Famaey \& Binney 2005), which is known to best match other galaxy
rotation curves (Gentile et al. 2011), finding that
$V_{\rm cor}$ becomes: 

\begin{equation}
V_{\rm cor}^2(r)=V_{\rm bar}^2(r)
\left(\frac{\sqrt{1+\frac{4 a_0 r}{V_{\rm bar}^2(r)}}+1}{2}\right),
\label{vmondnew1}
\end{equation}

\noindent
where $V_{\rm bar} $ is the Newtonian baryonic contribution to the rotation curve $\left(V_{\rm bar} =
\sqrt{V_{\rm stars}^2 +V_{\rm gas}^2} \right)$.

The scale-height of the gaseous disk is a
difficult quantity to determine in non edge-on galaxies: 
for dwarf galaxies, values ranging
roughly from 0.1 kpc to 0.5 kpc have been estimated (Elmegreen, Kim \&
Staveley-Smith 2001; Walter \& Brinks 2001).
We derived $V_{\rm gas}$ based on the HI surface density profile
(corrected for primordial He) and assumed a thickness of the
gaseous disk of 0.3 kpc. 

The contribution of the stars, $V_{\rm stars}$, was derived from the NIR
photometric profile, assuming a sech$^2$ distribution in the vertical
direction with a scale-height of $z_0 = h/5$, where $h$ is the
exponential scale-length (15.4 arcsec) that we derived from a fit to the
photometric profile. 

Kirby et al. (2008) gave a total stellar mass, which was based on the H-band
absolute luminosity (5.0 $\times$ 10$^7$ L$_{\odot}$) and with the
assumption of a mass-to-light (M/L)
ratio in the H-band of 
1.0. On the basis of the $B-H$ colour (3.16) published in Kirby et al., we found a
H-band M/L ratio of 0.66, using the model of Bell \& de Jong (2001). 
We considered the uncertainty in the M/L ratio to range from one third
of 0.66 up to two times 0.66
(de Jong \& Bell 2007, Bershady et al. 2011).

A MOND rotation curve fit to KK~246 is shown in Fig.
\ref{rc_all}. The distance was left as a free
parameter, but it was allowed to span only the 20\% uncertainty range
we assumed. The best-fit distance is 9.3 Mpc and the best-fit stellar
M/L ratio is at the lowest end of the permitted range (0.22). The
fit is overall excellent.
Because a distance of 9.3 Mpc is only marginally consistent with the
distances given in Karachentsev et al. (2006) and Tully et al. (2008),
we also show fits where the distance was fixed at 7.8 Mpc and 6.4
Mpc, respectively (Fig. \ref{rc_all}), where the fit quality gets
increasingly worse. For these smaller distances, the best-fit stellar
M/L ratios are 0.98 for a distance of 7.8 Mpc and 1.32 (the highest
end of the allowed range) for a distance
of 6.4 Mpc.

Alternatively, considerations similar to Angus et al. (2012) 
could be made: they varied the gaseous
scale-height while fitting the rotation curve of DDO 154, and found
that increasing the scale-height from 0.3 kpc to 0.7 kpc reduced the
$\chi^2$ by 30\%. The effect of increasing the scale-height of KK~246 would
mainly be to reduce the velocity of the MOND prediction at the innermost
point. 

Had we not applied the beam-smearing correction, but used 
the original Kreckel et al. (2011) rotation curve, little would have
changed: the
innermost point would have been slightly overestimated by the fits, the
quality of the fits would have been somewhat worse and the stellar M/L
ratio of the fit with a distance of 7.8 Mpc would have decreased to
0.40.

\begin{figure*}
\begin{center}
\includegraphics[width=1.0\textwidth]{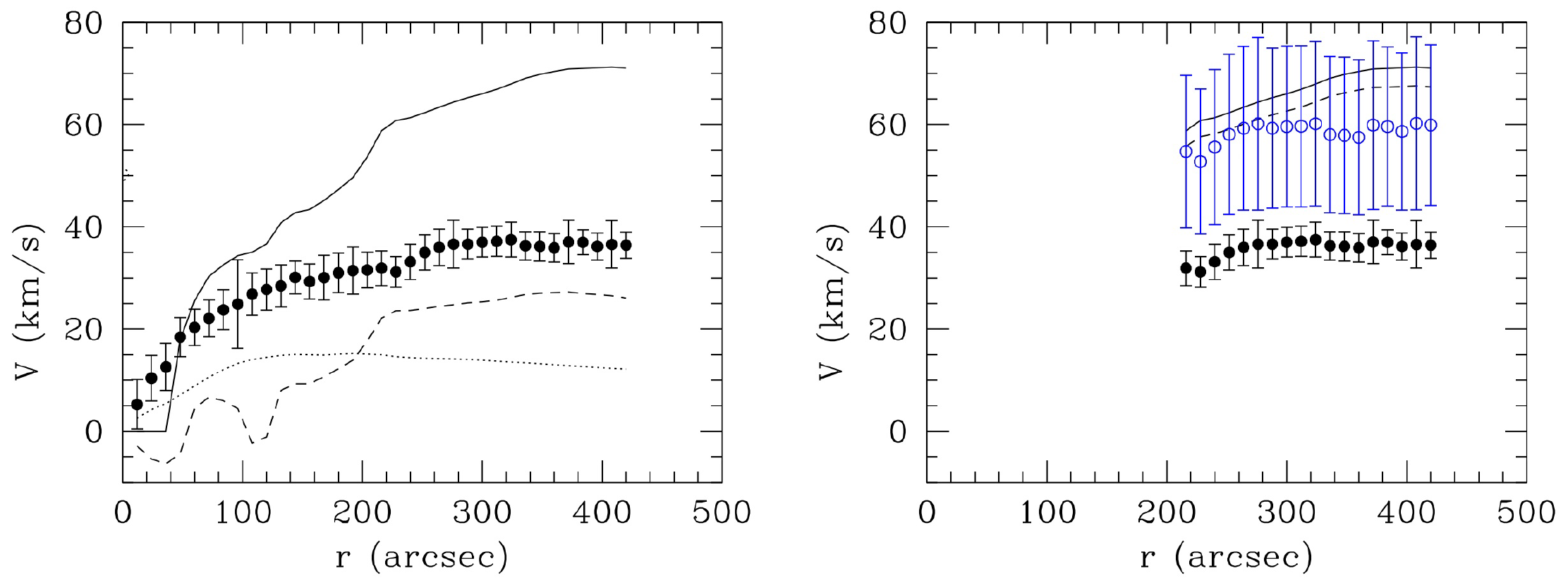}
\end{center}
\caption{Left: MOND mass model of Holmberg~II assuming the
rotation curve presented by Oh et al. (2011). The dotted line is the
contribution of the stellar disk assuming a M/L ratio in the 3.6$\mu$m
band of 0.39, the dashed line is the gas disk contribution, and the black
line is the MOND prediction using a distance of 3.4 Mpc (Karachentsev
et al. 2002). 
Right: rotation curve of Holmberg~II beyond 210 arcsec. 
The black points and the solid line are the same as in the right panel, and
the dashed line is the MOND prediction using a distance of 3.05 Mpc
(Hoessel et al. 1998).
The blue errorbars represent the range of outer
rotation velocities derived in the present paper and for the blue
  open circles we assumed a velocity that is at the middle of the allowed
  range.
} 
\label{oh_all}
\end{figure*}

\begin{figure*}
\begin{center}
\includegraphics[width=0.7\textwidth]{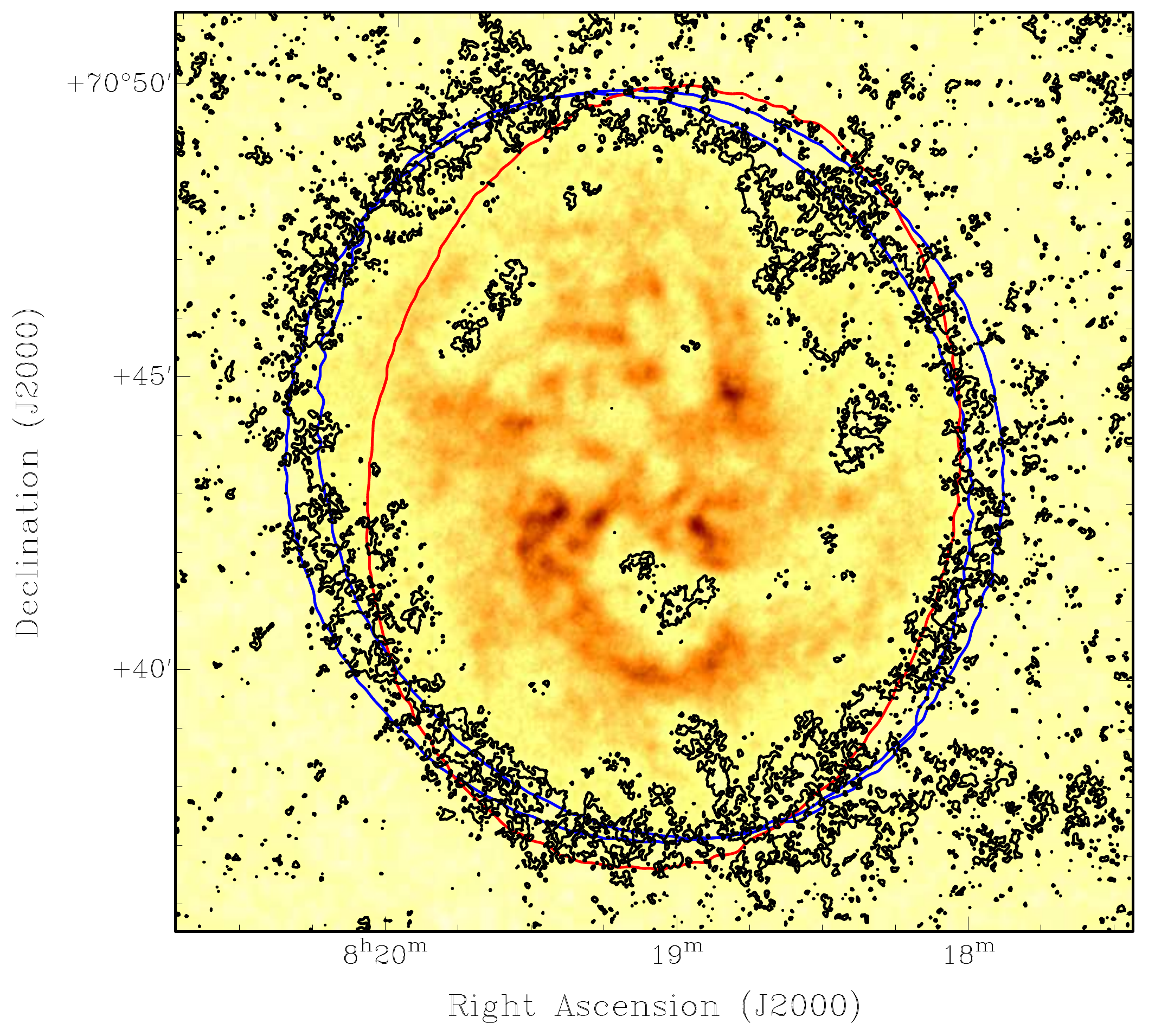}
\end{center}
\caption{Total HI map of Holmberg~II. The 1.5 $\times$ 10$^{20}$ atoms 
cm$^{-2}$ contour is shown for the observations (black), the model data cube
based on the parameters derived by Oh et al. (2011) (red), the
parameters derived in the present paper: the two blue lines represent
the two extremes of the outer inclination range allowed by the data
(20$^{\circ}$ to 35$^{\circ}$).
The FWHM beam size is 13.7 $\times$ 12.6 arcsec.
} 
\label{mom0_paper}
\end{figure*}

\section{Holmberg II}

In contrast to the isolated galaxy KK~246, Holmberg~II belongs to the M81 group of galaxies
(e.g. Karachentsev 2005, Walter et al. 2007). More precisely, Bureau
\& Carignan (2002) argue that it belongs to the subgroup dominated by
NGC 2403, the second largest galaxy of the M81 group.

The HI rotation curve of this galaxy, presented by Oh et al. (2011), is based on a
tilted-ring fit to the ``bulk velocity field'', which is a method for
deriving the velocity field by separating random motions from the bulk rotation
of a disk galaxy, and isolating only the latter (Oh et al. 2008). 
Oh et al. (2011) initially allowed all parameters to vary for each
ring, to search for the best values 
of centre, systemic velocity, position angle, and inclination; they
then performed a new tilted-ring fit with these parameters fixed 
to a smooth line going through the best-fit values, to look for the
best rotation curve. They found a rotation curve that remains
approximately flat (at a value slightly above $\sim$ 30 km s$^{-1}$) from 3 kpc to
the last measured radius of 10 kpc, even though for the mass models they
only considered data out to 7 kpc. 

As also discussed by Oh et al. (2011), the inclination is a concern in
Holmberg~II. In their tilted-ring fit, the ring-to-ring scatter in the
best-fit inclination values is quite high, and the outer inclination
that they found (relatively low, roughly between 40$^{\circ}$ and
50$^{\circ}$), raises some questions about the derived rotation
curve. 

We therefore investigated the inclination angle of Holmberg~II in some
more detail, especially in the outer parts of this galaxy. To this end, we made model
data cubes and compared them to the observed ones; in particular, we
compared the total HI maps, because they are most sensitive to the
choice of inclination angle. 
We made various models with different inclinations beyond a radius of
210 arcsec ($\sim$ 3.5 kpc when assuming a distance of 3.4 Mpc),
because this is where 
the discrepancy between the Oh et al. (2011) rotation curve and the one predicted by
MOND, regardless of the stellar M/L ratio, becomes significant
(Fig. \ref{oh_all}). To be consistent with Oh et al., we assumed a
scale-height of 0.28 kpc.

In Fig. \ref{mom0_paper}, we have compared 
the outer contours of the observed total HI map with three
models. The comparison shows that the parameters derived by Oh et
al. (2011) do not match the data well, and that their outer inclination angle
is too high. 
In contrast, a range of outer inclinations between 20$^{\circ}$ and
35$^{\circ}$ that are consistent with the one ($\sim$25$^{\circ}$) based on the
baryonic Tully-Fisher relation, provides a closer match to the observed
total HI map. We also shifted the position of the centre 
by $\sim$ 25 arcsec (about two beam FWHM) towards the NE and we slightly
changed the outer position angle to match the outer contours of the
total HI map. The better match of the present model is also due to a
different position angle around a radius of 6 arcmin, which is likely
to be caused by
a non-circular component in the velocity that causes a shift between
the kinematical and morphological position angles. However, owing to the
irregular shape of the total HI emission in the innermost part (and, to a
lesser extent, in the outermost parts),  
an accurate determination of
the radial dependence of the morphological position angle is unfeasible.

Correcting the (asymmetric-drift corrected) observed rotation 
curve for an outer inclination between 20$^{\circ}$ 
and 35$^{\circ}$ gives an outer
rotation velocity range between 40 km s$^{-1}$ and 75 km s$^{-1}$
(Fig. \ref{oh_all}). 
As errorbars, we considered the squared sum of (1) the original
uncertainty given by Oh et al. (2011), corrected for the
inclination, and (2) the difference between the velocity using an
inclination of 20$^{\circ}$ and the velocity using an inclination of
35$^{\circ}$.  
This implies that, including these
systematic errors in the inclination angle, the outer rotation curve
is compatible
with both the MOND prediction and 
the baryonic Tully-Fisher relation. 
We compared the MOND prediction using two accurate derivations of the
distance: 3.05 Mpc by Hoessel et al. (1998) using Cepheids, and 3.4
Mpc by Karachentsev et al. (2002) using the tip of the red giant
branch method. Fig. \ref{oh_all} shows that with the new
inclination angles in the outer part, the rotation curve of Holmberg
II is consistent with the MOND prediction using both estimates of the distance.
A reliable derivation of the inclination
variation over the full range of radii is a very challenging task,
because in such a low-inclination dwarf spiral galaxy 
the shape of the HI emission in both the single channel maps and the
total HI map at smaller radii is dominated
by shells and holes, thus it goes beyond the goal of the present
paper.
We note that the errorbars in a rotation curve that includes 
systematic errors in the inclination are larger than the
point-to-point scatter, but reflect the true uncertainty in each
velocity point.
Additionally, if we assume that there is an
extreme flare in the outer parts, increasing the scale-height up to 1
kpc, the inclination range changes to 22$^{\circ}$-39$^{\circ}$, and
the corresponding velocity range changes to 36-68 km s$^{-1}$, which
is still
consistent (though marginally) with the MOND prediction.

\section{Conclusions}
\label{Conclusions.sec}

We have investigated the kinematics of two dwarf spiral
galaxies within the framework of MOND. These two galaxies are in very
different environments: KK~246 is in the Local Void (Tully et
al. 2008), whereas Holmberg II is a member of the M81 group of galaxies
(Karachentsev 2005, Walter et al. 2007). 

For KK~246, we used the HI rotation curve presented in Kreckel et
al. (2011) and the H-band photometry obtained by Kirby et
al. (2008). We have found that, although the best-fit distance is slightly
large, within its uncertainties those of the stellar M/L ratio, the
observed rotation curve is consistent
with the MOND prediction. 

The other galaxy, Holmberg II, is at first sight inconsistent with
MOND, using the rotation curve of Oh et al. (2011). However, after modelling
the HI data cube and comparing the observed and modelled total HI
maps, we found that the inclination angle in the outer parts had been
overestimated. The inclination angle is indeed closer to face-on than
had previously been determined. This results in a rotation curve
with a higher amplitude and a larger uncertainty; 
this new rotation curve is
compatible with the MOND prediction. In addition, with the new inclination, 
Holmberg II now falls on the baryonic Tully-Fisher relation (its baryonic
Tully-Fisher velocity is $\sim$ 65 km s$^{-1}$). 
It is clear that the limited resolution of the observations, and the
inherent uncertainties in the geometrical parameters of these galaxies
limits their application to either the confirmation or falsification of dark
matter or alternative theories. 
Further decreasing systematic uncertainties in the determination of
rotation curves of dwarf galaxies residing in different environments
will thus be very useful in the future in testing the
MOND paradigm more rigourously in the realm where it is supposed to work best. 

\section*{Acknowledgements}

We thank the referee for useful comments that improved the quality of
this paper.
GG is a postdoctoral researcher of the FWO-Vlaanderen (Belgium). BF
acknowledges the support of the AvH foundation. Some of SHO research
was carried at the ``Centre for All-sky Astrophysics'', which is an Australian
Research Council Centre of Excellence, funded by grant CE11E0090.

\label{lastpage}

\end{document}